\documentclass[10pt,conference]{IEEEtran}
\pdfoutput=1 % Force PDFLATEX on ARXIV
\usepackage{cite}
\usepackage{amsmath,amssymb,amsfonts}
\usepackage{graphicx}
\usepackage{textcomp} % For trademark symbol
\usepackage{xcolor}
\usepackage{multicol} % Put two equations side by side

\def\BibTeX{{\rm B\kern-.05em{\sc i\kern-.025em b}\kern-.08em
    T\kern-.1667em\lower.7ex\hbox{E}\kern-.125emX}}
    
\usepackage[utf8]{inputenc}
\usepackage{xcolor}
\usepackage[xindy,style=index,numberedsection=false]{glossaries}
\usepackage{url}
\usepackage{multirow}
\usepackage[]{algpseudocode,algorithm}
\algtext*{EndWhile}% Remove "end while" text
\algtext*{EndIf}% Remove "end if" text
\algtext*{EndFor}% Remove "end if" text
\algtext*{EndFunction}% Remove "end if" text
\usepackage{textcomp}
\usepackage[noabbrev]{cleveref}
\usepackage{balance}
\usepackage{graphicx}
\usepackage{subcaption}
\usepackage{booktabs}
\usepackage{makecell}
\usepackage{caption}
\usepackage{siunitx}
\usepackage{tikz}
\usepackage{circledsteps}

\usepackage{setspace}
\let\Algorithm\algorithm
\renewcommand\algorithm[1][]{\Algorithm[#1]\setstretch{1.0}}
\algnewcommand{\LineComment}[1]{\State \(\triangleright\) #1}

\crefname{algocf}{alg.}{alg.}
\Crefname{algocf}{Alg.}{Alg.}
\crefname{lstlisting}{listing}{listings}
\Crefname{lstlisting}{Listing}{Listings}
\crefname{algorithm}{alg.}{alg.}
\Crefname{algorithm}{Alg.}{Alg.}

\AtBeginDocument{} % Used in double-equations on one line;

\newcommand{\ra}[1]{\renewcommand{\arraystretch}{#1}}
\makeatletter
\newcommand{\linebreakand}{%
  \end{@IEEEauthorhalign}
  \hfill\mbox{}\par
  \mbox{}\hfill\begin{@IEEEauthorhalign}
}
\makeatother

\newacronym{nlp}{NLP}{Natural Language Processing}
\newacronym{fpga}{FPGA}{Field-Programmable Gate Array}
\newacronym{spmv}{SpMV}{Sparse matrix-vector multiplication}
\newacronym{dsl}{DSL}{Domain-Specific Language}
\newacronym{csc}{CSC}{Compressed Sparse Column}
\newacronym{csr}{CSR}{Compressed Sparse Row}
\newacronym{raw}{RAW}{Read-After-Write}

\newacronym{gpu}{GPU}{Graphics Processing Unit}
\newacronym{vm}{VM}{Virtual Machine}
\newacronym{ir}{IR}{Intermediate Representation}
\newacronym{sp}{SP}{Stream Processor}
\newacronym{sm}{SM}{Stream Multiprocessor}
\newacronym{jvm}{JVM}{Java Virtual Machine}
\newacronym{pcie}{PCIe}{PCI Express}
\newacronym{ipc}{IPC}{Instructions per cycle}
\newacronym{dag}{DAG}{Directed Acyclic Graph}
\newacronym{dl}{DL}{Deep Learning}
\newacronym{um}{UM}{Unified Memory}
\newacronym{fifo}{FIFO}{First In First Out}
\newacronym{nidl}{NIDL}{Native Interface Definition Language}
\newacronym{nfi}{NFI}{Native Function Interface}
    
\begin{document}

\title{DAG-based Scheduling with Resource Sharing for Multi-task Applications in a Polyglot GPU Runtime}

\author{\IEEEauthorblockN{Alberto Parravicini}
\IEEEauthorblockA{\textit{Politecnico di Milano} \\
Milan, Italy \\
alberto.parravicini@polimi.it}
\and
\IEEEauthorblockN{Arnaud Delamare}
\IEEEauthorblockA{\textit{Oracle Labs} \\
Z\"{u}rich, Switzerland \\
arnaud.d.delamare@oracle.com}
\and
\IEEEauthorblockN{Marco Arnaboldi}
\IEEEauthorblockA{\textit{Oracle Labs} \\
Z\"{u}rich, Switzerland \\
marco.arnaboldi@oracle.com}
% \linebreakand
\and
% \IEEEauthorblockN{Lukas Stadler}
% \IEEEauthorblockA{\textit{Oracle Labs} \\
% Linz, Austria \\
% lukas.stadler@oracle.com}
% \and
% \IEEEauthorblockN{Rene Mueller}
% \IEEEauthorblockA{\textit{Nvidia} \\
% Z\"{u}rich, Switzerland \\
% muellren@acm.com}
% \and
\IEEEauthorblockN{Marco D. Santambrogio}
\IEEEauthorblockA{\textit{Politecnico di Milano} \\
Milan, Italy \\
marco.santambrogio@polimi.it}
}

\maketitle

\begin{abstract}
GPUs are readily available in cloud computing and personal devices, but their use for data processing acceleration has been slowed down by their limited integration with common programming languages such as Python or Java. Moreover, using GPUs to their full capabilities requires expert knowledge of asynchronous programming.
In this work, we present a novel GPU run time scheduler for multi-task GPU computations that transparently provides asynchronous execution, space-sharing, and transfer-computation overlap without requiring in advance any information about the program dependency structure.
We leverage the GrCUDA polyglot API to integrate our scheduler with multiple high-level languages and provide a platform for fast prototyping and easy GPU acceleration. We validate our work on 6 benchmarks created to evaluate task-parallelism and show an average of 44\% speedup against synchronous execution, with no execution time slowdown compared to hand-optimized host code written using the C++ CUDA Graphs API.
\end{abstract}


\glsresetall
\section{Introduction}\label{sec:intro}

\glspl{gpu} are often heralded as the optimal solution to achieve extremely high throughputs in domains such as Deep Learning, financial simulations, and graph analytics. Even if \glspl{gpu} are made available by most cloud providers and are commonly found in personal devices, using \glspl{gpu} for data processing acceleration has been hampered by their limited integration with high-level programming languages such as Python or Java. Fully exploiting the hardware resources of \glspl{gpu} requires a deep understanding of their architecture and creates a steep learning curve that takes a long time for programmers to overcome. As a consequence, the adoption of \glspl{gpu} is often limited to specific domains for which libraries or \glspl{dsl} that abstract and mask the \gls{gpu} computation are available. 

% This paragraph might change if we present the whole GrCUDA.
On the other hand, APIs offering complete control over the \gls{gpu} still require efforts to unleash the full hardware potential, e.g. to overlap and synchronize multiple computations, or to overlap computation with data transfer from and to the \gls{gpu}. In this work, we present a novel \gls{gpu} runtime scheduler that transparently provides all these optimizations without requiring in advance any information about the structure of the computation. We leverage the Graal polyglot \gls{vm}
% \footnote{\url{https://github.com/oracle/graal}}
\cite{wurthinger2013one, wimmer2012truffle,duboscq2013graal}, and the GrCUDA environment \cite{grcuda}, to run GPU kernels from languages such as Python, Java, and Ruby and obtain full control over the \gls{gpu} runtime to optimize scheduling, data transfer, and execution. 
We target the CUDA platform, but similar considerations would work for OpenCL \cite{stone2010opencl}, given the availability of a managed execution environment.

% As the amount of data in today's analytical workloads grows much faster than the computational capabilities offered by traditional computing platforms, and state-of-the-art data processing and machine learning techniques become more complex and computationally demanding, it is required to look at alternatives to traditional CPUs, and consider modern hardware accelerators such as \glspl{gpu} and heterogeneous computing platforms.

\subsection{Motivation}

Modern \glspl{gpu} allow multiple computations to run asynchronously and concurrently to leverage task-level parallelism: experienced programmers accelerate their programs by prefetching and overlapping data transfers with computations on different data or even take advantage of hardware space-sharing by overlapping multiple independent computations.

\begin{figure}[t]
    \centering
    \includegraphics[width=\columnwidth]{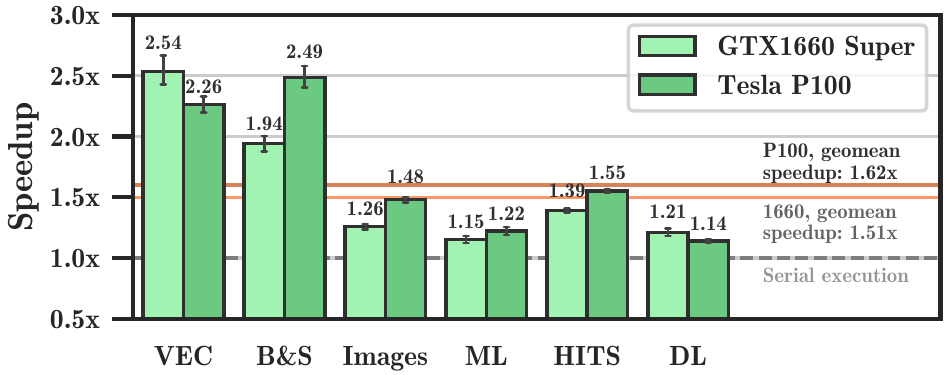}\\
    \caption{Achievable speedup in C++ CUDA with hand-tuned GPU data transfer and execution overlap. Fine-tuned space-sharing and execution-transfer overlaps can accelerate GPU computations by more than 50\%.}
    \label{fig:cuda_speedup_example}
\end{figure}

\begin{figure}[t]
    \centering
    \includegraphics[width=\columnwidth]{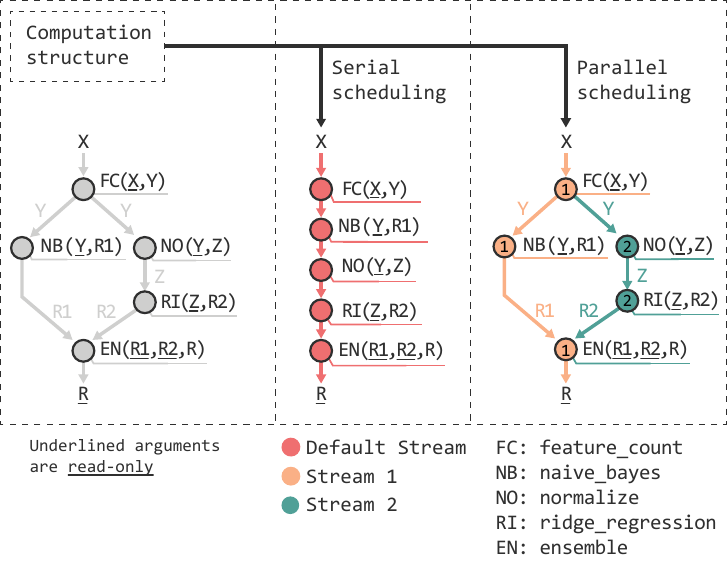}\\
    \caption{The two branches in this computation are independent, and can be scheduled and executed in parallel. Edges are labeled with the argument that cause a data dependency.}
    \label{fig:schedule_example}
\end{figure}

\Cref{fig:cuda_speedup_example} shows how much speedup can be extracted in different CUDA benchmarks (presented in \cref{sec:benchmarks}) by hand-crafting these optimizations.
Computations with opportunities for task-level parallelism are common: even the simple machine learning pipeline in \Cref{fig:schedule_example} has two independent branches whose results are combined at the end. Instead of executing computations sequentially, a skilled programmer
% understands which tasks can run in parallel and schedule them on different execution streams.
schedules independent tasks on separate execution streams.
However, achieving full utilization of the \gls{gpu} often requires extensive debugging and profiling, even by experienced users \cite{luitjens2015cuda}; performance may also depend on the data available at run time, and it cannot be perfectly fine-tuned by the programmers.

Our work aims to provide a low-profile runtime that can automatically leverage untapped \gls{gpu} resources in multi-task computations to provide speedups identical to what a skilled programmer can achieve by hand, lowering the barrier of access to \glspl{gpu} with no performance compromises.

\subsection{Contributions}
In this work, we present a novel low-profile run time scheduler for multi-task and asynchronous \gls{gpu} computations. 
 
Our scheduler automatically infers data dependencies between \gls{gpu} kernels, models them using a \gls{dag}, and enable asynchronous CPU and \gls{gpu} execution without users having to define any synchronization event or dependencies manually. More importantly, dependencies and scheduling are computed entirely at run time, without defining the computation structure in advance, and without constraints on the host language control flow.
 
This work is implemented as an extension of GrCUDA, a polyglot CUDA API based on GraalVM\cite{wurthinger2013one}. GrCUDA is implemented as a Truffle \gls{dsl} \cite{wimmer2012truffle} and provides access to \gls{gpu} acceleration to languages supported by GraalVM, such as Java, Scala, JavaScript, R, and Python.
Our scheduler is available in all these languages; moreover, any new feature or optimization to our scheduler will be available without language-specific modifications. Our scheduler enables GrCUDA to become a valid solution for general-purpose \gls{gpu} acceleration, focusing on fast prototyping and integration with high-level languages that do not currently have a strong \gls{gpu} support.
 
We evaluate our scheduler on 6 benchmarks from different domains that exhibit opportunities for task-level parallelism and show an average of 44\% speedup against the serial GrCUDA scheduler and no significant slowdown against hand-optimized scheduling written using the C++ CUDA Graphs API; finally, we analyze hardware utilization to understand how well each benchmark can exploit data transfer-computation overlap and untapped \gls{gpu} resources. The source code for our scheduler and benchmarks is openly available\footnote{\url{github.com/AlbertoParravicini/grcuda}}.

In summary, we make the following contributions:
\begin{itemize}
    \item A run time scheduler based on GrCUDA that automatically infers dependencies between \gls{gpu} computations and dynamically schedules them to maximize transfer-computation overlap and space-sharing (\cref{sec:implementation}).
    \item A suite of 6 benchmarks to evaluate \gls{gpu} asynchronous computations, task-parallelism, and space-sharing hardware utilization (\cref{sec:benchmarks,sec:hardware_metrics}).
    \item An evaluation of how our scheduler provides an average of 44\% speedup against the serial GrCUDA scheduler and no slowdown against hand-optimized scheduling based on the CUDA Graphs API (\cref{sec:grcuda_speedup,sec:grcuda_cuda_speedup}).
\end{itemize}

\section{Related Work}\label{sec:soa}

Expressing a multi-task computation as a \gls{dag} that can be used to estimate an effective scheduling and to optimize execution is a concept that has already been explored with a great degree of success: well-known work that covers \gls{gpu} computations includes Nvidia's CUDA Graphs\cite{cudagraphapi} and Google's Tensorflow \cite{abadi2016tensorflow}; academic research has also shown interest in domains such as distributed and heterogeneous computing \cite{marchal2018parallel, xu2020dag, ozkaya2019scalable},
% Domain-Specific Architectures (with the Pixel Visual Core \cite{10.5555/77493})
and presented valuable theoretical results \cite{mayer2017tensorflow, marchetti2020complexity}.
CUDA Graphs are a programming model recently released by Nvidia used to define a \gls{dag} of inter-dependent computations and execute them asynchronously. Computations between dependencies must be specified manually using CUDA events, or with a fairly complex custom API \cite{cudagraphapi}. CUDA Graphs also present initialization overheads due to graph creation \cite{cudagraph}.
Instead, TensorFlow allows users to express a \gls{dag} of computations through a \gls{dsl} embedded in languages such as Python. TensorFlow is mostly intended for \gls{dl}; expressing custom kernels for other domains, while supported, is not straightforward and requires significant manual integration effort. We do not deem a direct comparison with our work to be meaningful, as TensorFlow presents design choices specifically targeted towards \gls{dl}. 

Indeed, the most common approach to \gls{dag}-based scheduling, as seen in CUDA Graphs and TensorFlow, is to specify the program flow in advance: this choice simplifies the computation of an optimal scheduling and amortizes overheads in case of repeated computations, and is certainly suitable for specific domains such as \gls{dl}. On the other hand, the approach presented in this work is to capture \gls{gpu} computations through a low-profile runtime, without the need to specify dependencies manually. As such, we do not impose any limitation over the host program control-flow (e.g. conditional statements, function calls, recursion, library calls) so that the program flow can change over time. In the simplest scenario, a programmer might use multiple kernel implementations optimized for different input sizes or use different pre-processing procedures based on the input data language; selecting the appropriate kernel is done simply through conditional statements in the host language (e.g. a switch-case in Python), without requiring custom APIs or defining multiple \glspl{dag} in advance.

Computing dependencies at runtime on heterogeneous architectures has been explored by the XKaapi runtime \cite{gautier2013xkaapi}. Contrary to our approach, XKaapi uses a work-stealing strategy that computes dependencies every time an idle thread looks for a task to execute. It handles \gls{gpu} memory as a queue of blocks instead of leveraging the fine-grained flexibility of \gls{um}, and its complex API is limited to C++.

Many techniques to virtualize \glspl{gpu} usage have emerged recently, although space-sharing is still considered an open challenge \cite{hong2017gpu}. Ravi et al. \cite{ravi2011supporting} consolidate kernels from different \glspl{vm} to increase utilization. TornadoVM \cite{fumero2019dynamic}, which is also compatible with GraalVM, translates annotated Java code for heterogeneous hardware; it profiles the code to understand the most suitable backend, and infers automatically data-dependencies and data-transfer through a task graph obtained from the computations specified in advance by the user \cite{clarkson2018exploiting}.

Today, most programming languages have ways to achieve \gls{gpu} acceleration, including Python (PyCUDA, PyOpenCL \cite{kloeckner_pycuda_2012}), Java
% (Jcuda\footnote{\label{jcuda}Jcuda: \url{www.jcuda.org}, GPU.js: \url{https://gpu.rocks}})
(Jcuda\footnote{Jcuda: \url{www.jcuda.org}, GPU.js: \url{https://gpu.rocks}})
and JavaScript
% (GPU.js\footnotemark[\ref{jcuda}]).
(GPU.js).
However, none of these libraries can automatically handle asynchronous computations and space-sharing. Moreover, each library has different APIs, supported features, and update cycles, greatly limiting code portability and interoperability; instead, the unified GrCUDA runtime and API immediately provides each new feature (such as our scheduler) to all languages supported by GraalVM without any change in the host code.

Recent research on \gls{gpu} space-sharing includes the work of Wen et al. \cite{wen2017merge}, who showed how concurrent execution of some \gls{gpu} kernels without data dependencies delivers up to 1.5x speedup; Qiao et al. \cite{qiao2020unveiling} uses concurrent kernel execution to achieve 2.5x speedup over serial execution on multiresolution image filters. 
% Other works (Baymax \cite{chen2016baymax}, Effisha \cite{chen2017effisha}) focus on multi-application space sharing. 
Baymax \cite{chen2016baymax} focuses on multi-application space sharing. 
DCUDA \cite{guo2019dcuda} provides scheduling for different applications on multiple \glspl{gpu}, and shows how CUDA \gls{um} causes an average slowdown below 1\%. We focus on single-application space-sharing, i.e. applications composed of many kernels that can run in parallel, but considerations on hardware utilization 
% and techniques for low-profile scheduling 
are valid in both single and multi-application space-sharing.
As GrCUDA seamlessly integrates with CUDA, our work directly benefits from improvements on the CUDA API and drivers. For example, we could leverage alternative \gls{um} implementations optimized to overlap data transfer with computations, such as HUM \cite{jung2020overlapping}.
\section{Background}\label{sec:context}

% \glspl{gpu} are specialized computer architectures conceived to process a large amount of data in parallel by running computational kernels on each data item (e.g. pixels the of an image), or on small chunks of data items.
% The architecture of a \gls{gpu} has multiple levels: \glspl{gpu} manufactured by Nvidia divide the computation across \glspl{sm}, each containing multiple \glspl{sp} (or CUDA cores) that execute the same kernel on different data items. 
% \gls{gpu} execution usually happens asynchronously with respect to the CPU, which can perform other computations in the meantime. This programming model enables heterogeneous CPU-GPU computing but requires synchronization points to guarantee correctness and data consistency. 
% To simplify memory management, CUDA supports \acrfull{um}, a single memory address space accessible by both CPU and \gls{gpu} without the need for manual data movements. \gls{um} removes some tedious operations, but it has some drawbacks: on \gls{gpu} up to the Maxwell architecture, CPU accesses to \gls{um} produce a run time error if a \gls{gpu} kernel is running. The page migration mechanism introduced with Pascal lifted this limitation; a good amount of care is still required to avoid data hazards and performance issues caused by a sub-optimal transfer of memory pages.

\glspl{gpu} allow the execution of multiple kernels at the same time. If enough resources are available - e.g. free \gls{sm}, the main \gls{gpu} computation units - the \gls{gpu} will perform space-sharing and run kernels in parallel. Modern \glspl{gpu} often have enough resources to perform space-sharing without significantly degrading the performance of individual kernels \cite{qiao2020unveiling}: space-sharing can improve the occupancy of \glspl{sm} and exploit under-utilized hardware resources.
Kernels run in parallel and asynchronously thanks to CUDA streams: kernels are executed in issue-order on a stream, but different streams proceed independently. Space-sharing is key for better \gls{gpu} resource usage,
% and greater opportunities for asynchronous execution
but requires even greater care to ensure that programs run correctly. Synchronizing a single stream through the \texttt{cudaStreamSynchronize} function blocks the host execution and is generally acceptable only if the host requires the output of a \gls{gpu} kernel. A more flexible approach is the use of CUDA events\footnote{\url{ docs.nvidia.com/cuda/cuda-runtime-api/group__CUDART__EVENT.html}}, which allow streams to synchronize with each other without blocking the host execution. Using CUDA events to efficiently synchronize multiple complex streams by hand can be cumbersome:
% , and is often cause for programming mistakes:
instead, our solution masks CUDA events and greatly simplifies the optimal scheduling of asynchronous computations (\cref{sec:implementation}).

% Placed here for formatting
\begin{figure}[t]
    \centering
    \includegraphics[width=\linewidth]{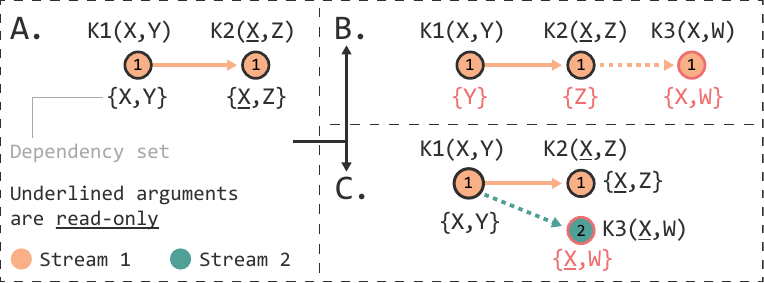}\\
    \caption{Dependency computations with read-only arguments. Updates to the DAG and to dependency sets are highlighted.}
    \label{fig:dag_dependency_read}
\end{figure}

In CUDA, computations are divided in \textit{blocks}, each composed of an equal number of \textit{threads} (from 32 to 1024). Bigger blocks imply fewer blocks running concurrently, but more possibilities of sharing data through fast block-wide shared-memory. 
Users can choose the number of threads per block; depending on the kernel implementation, users can also choose the number of blocks.
While a small number of blocks increases space-sharing, achieving a performance improvement depends on the characteristics of the kernels.
% As it is common nowadays, we choose the number of blocks for kernels in our benchmark suite based on performance, and rely on grid-stride loops to process arbitrarily large input data.
These considerations also apply to OpenCL: \textit{streams} are called \textit{command queues}, \textit{threads} are \textit{work items}, and the event model is similar to CUDA.
Streams can also improve performance by overlapping asynchronous data transfer (from CPU to \gls{gpu} or vice-versa) and computations: in \Cref{fig:schedule_example} the transfer of array \texttt{r1} (required by kernel \texttt{NB}) can be overlapped with execution of kernel \texttt{NO}. This approach is fruitful with repeated execution of simple kernels on different data batches, as the data transfer takes a significant amount of the total execution time. When using \gls{um} on \glspl{gpu} that offer page migration, it is beneficial to prefetch data instead of relying on migrations caused by page faults: our scheduler can prefetch data automatically, reducing the burden of \gls{gpu} optimization.

We leverage GraalVM, a Java \gls{vm}
able to run and combine languages that compile to Java bytecode (e.g. Scala) and custom implementations of other languages such as JavaScript, R, and Python \cite{wurthinger2013one}.
This interoperability is possible thanks to the Truffle Abstract Syntax Tree interpreter
\cite{wimmer2012truffle}, which guarantees high-performance through partial evaluation of repeated portions of code.
Our work is an extension of GrCUDA, a CUDA language binding implemented as a Truffle \gls{dsl}. GrCUDA can be seen as a polyglot API, as it provides \gls{gpu} acceleration to all languages supported by GraalVM.
\section{Scheduler Design Methodology}\label{sec:implementation}

This section details the design and implementation of our scheduler. We provide a definition of our computation \gls{dag} (\cref{sec:dag}). Then, we define our scheduler's architecture and its integration with the CUDA runtime (\cref{sec:architecture,sec:policies}). Finally, we show how to leverage the language design of GrCUDA to provide asynchronous \gls{gpu} computation without any detriment to accessibility (\cref{sec:language}).

\subsection{Computation DAG and Dependency Sets}\label{sec:dag}

The cornerstone of our scheduler is a Computation \gls{dag} that represents relationships between computations that involve the \gls{gpu}. Vertices of the \gls{dag} are \textit{computational elements}: \gls{gpu} kernels, memory accesses by the CPU host program to GrCUDA \gls{um}-backed arrays, and pre-registered or user-defined library functions such as RAPIDS\footnote{\url{developer.nvidia.com/rapids}}. Using GrCUDA and GraalVM, each element can be encapsulated through an object to keep track of its state. In the case of kernels, the object tracks its configuration (e.g. the number of blocks), its input arguments, and if the computation is active.

Edges of the \gls{dag} are \textit{data dependencies} between computational elements. Dependencies are inferred automatically instead of being manually specified by the user through handles or other APIs. Inferring data dependencies is possible as GrCUDA uses a managed execution environment that allows object encapsulation of inputs, removing the risk of pointer aliasing typical of native languages (e.g. having multiple pointers referring to the same memory area). The scheduler employs data dependencies modeled through the \gls{dag} to associate computational elements to CUDA streams and introduces synchronization events if required. To compute dependencies, we associate with each computational element a \textit{dependency set}. This set initially contains all arguments of the computational element. An argument in the set is removed when a subsequent computation uses and modifies the same argument, defining a data dependency on it; once a set is empty, the corresponding computational element can no longer introduce dependencies. 
Read-only kernel arguments (specified as in \cref{sec:language}) can be treated with special rules to avoid adding unnecessary dependencies to the \gls{dag}. If possible, they will be ignored in the dependency computations: for example, if two kernels use the same read-only input array, they will be executed concurrently on different streams. \Cref{fig:dag_dependency_read} shows a kernel that modifies an argument and is followed by another kernel that uses the same argument as read-only \Circled{A}. If a third kernel with the same input is added, it will depend on the second kernel if it modifies the argument (a write-after-read anti-dependency) \Circled{B}, and it will depend on the first kernel if it uses the argument as read-only \Circled{C}; it will not, however, depend on both kernels. In case \Circled{C}, the read-only argument adds a new dependency through \texttt{X}, but the \textit{dependency set} of the parent kernel \texttt{K1} is not updated: if a new kernel requires \texttt{X} as read-only argument, it will depend on \texttt{K1}, otherwise it will depend on both \texttt{K2} and \texttt{K3}, and all \textit{dependency sets} will be updated.

\begin{figure}[t]
    \centering
    \includegraphics[width=\linewidth]{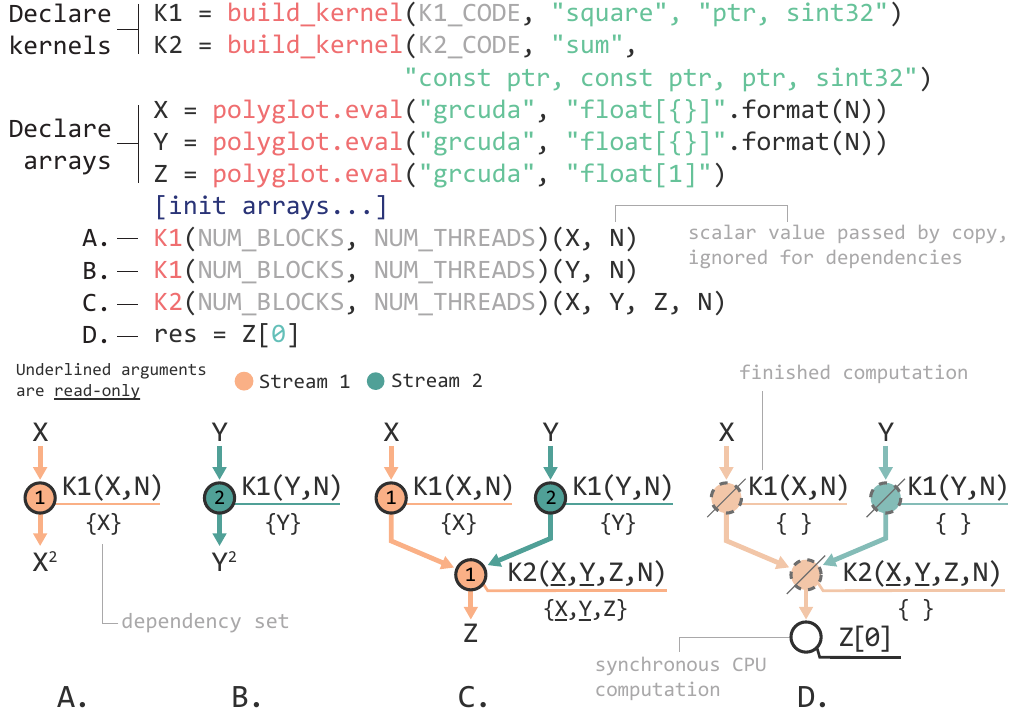}\\
    \caption{Example of scheduling for the \textbf{VEC} benchmark. We show both the GrCUDA code and the resulting DAG.}
    \label{fig:code_schedule}
\end{figure}

The \gls{dag} is built at run time, not at compile-time or eagerly. Users do not have to worry about their host program's control flow, as we dynamically add and schedule new computations as users provide them. 
% While computations in \cref{fig:benchmark_structure} are presented as complete \glspl{dag} for clarity, our scheduler does not know the full \textit{logical} \gls{dag} structure of a given program:
Our scheduler is unaware of the full \textit{logical} \gls{dag} structure of a given program, although we usually show complete \glspl{dag} for clarity (e.g. \Cref{fig:schedule_example,fig:benchmark_structure}).
Instead, the scheduler updates the current graph frontier, i.e. the computations that are currently active. 
% This choice is critical to allow users not to define in advance the structure of their program 
This choice is key to enable the dynamic creation of the \gls{dag} 
and does not introduce limitations on the optimizations that our scheduler could perform. 
We track each kernel's historical performance and scheduling to allow the creation of heuristics that guide future scheduling of the same kernel.
% For example, the scheduler could decide whether it is more beneficial to prefetch the kernel's input data or to rely on the \gls{gpu} page fault mechanism.

In GrCUDA, arrays are backed by \gls{um}, which simplifies data movement and accesses from the CPU without large performance penalties \cite{guo2019dcuda}. As the CPU can schedule accesses to these arrays at any point (even while the \gls{gpu} is running), we model these accesses as \textit{computational elements}. If the access introduces a data dependency on a \gls{gpu} computation, the scheduler ensures that the CPU waits for that computation to end.
To keep overheads as low as possible, array accesses that do not introduce data dependencies with respect to \gls{gpu} kernels are executed immediately, without modeling them as \gls{dag} elements: this is the case of consecutive accesses or accesses performed while no \gls{gpu} computation is active.
Pre-registered libraries can also take advantage of our scheduler if they expose the choice of execution stream in their API. If not, they are scheduled synchronously to guarantee correctness.

\Cref{fig:code_schedule} shows the GrCUDA code (with Python as host) of the \textbf{VEC} benchmark (\cref{sec:benchmarks}). For each kernel invocation in the host, the scheduler adds a computational element to the \gls{dag}, updates the dependency sets of active computations, and provides a CUDA stream for execution. Executing \texttt{K2} (\Cref{fig:code_schedule}, \Circled{C}) requires a CUDA event to ensure that \texttt{K1} is completed. Accessing \texttt{Z} on the CPU ensures that all computations are completed: they will no longer contribute to new dependencies. The host code does not need to care about the scheduling, and it can be written as if it were run sequentially, with no explicit mention of synchronization points or streams.

\subsection{System Architecture}\label{sec:architecture}

\begin{figure}[t]
    \centering
    \includegraphics[width=\columnwidth]{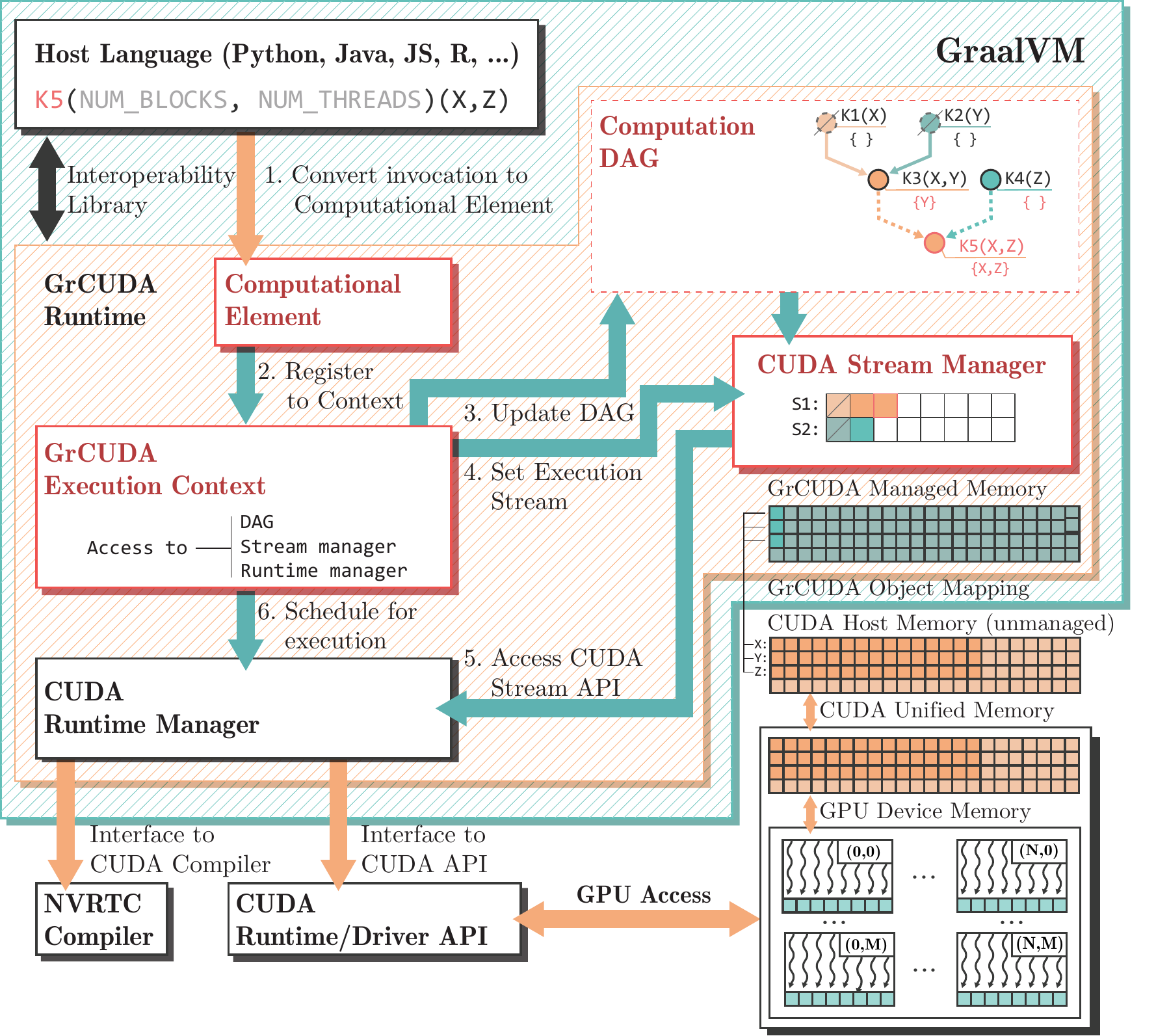}\\
    \caption{Our scheduler integrates with the existing GrCUDA architecture. New components are highlighted in red.}
    \label{fig:architecture}
\end{figure}

\Cref{fig:architecture} shows the main components of our scheduler and their integration with the existing GrCUDA architecture.
The \textit{\gls{gpu} execution context} tracks declarations and invocations of \gls{gpu} computational elements \Circled{1}. When a new computation is created or called, it notifies the execution context \Circled{2} so that it updates the \gls{dag} with  data dependencies of the new computation \Circled{3}.
The \gls{gpu} execution context uses the DAG to understand if the new computation can start immediately or if it must wait for other computations to finish. Computations are overlapped using different CUDA streams, assigned by the \textit{stream manager} based on dependencies and free resources (\cref{sec:policies}) \Circled{4}.
The \textit{stream manager} \Circled{5} and the \textit{execution context} \Circled{6} interact with the \gls{gpu} through an intermediate layer that exposes the CUDA API to GrCUDA.

\gls{gpu} computations are asynchronous and do not require synchronization against previous computations on the stream where they are executed, as CUDA guarantees sequential execution of computations scheduled on a stream. We synchronize streams with CUDA events without blocking the host CPU. Each computation is associated with an event to provide a precise synchronization point instead of blocking the entire stream. If the CPU requires data for a computation, we synchronize only the streams that are currently operating on this data. The scheduler considers \gls{gpu} computations as active until the CPU requires their result or one of their children.
We use active computations to identify dependencies (unless their \textit{dependency set} is empty) and empty streams.

\subsection{Scheduling Policies and Stream Management}\label{sec:policies}
 
A scheduler is \textit{serial} if computations are executed one after the other in the order defined by the user, and their execution or data transfer do not overlap. A \textit{parallel} scheduler allows overlaps to happen, and data dependencies determine the order in which computations are executed. In a \textit{synchronous} scheduler the CPU host program waits for \gls{gpu} computations to finish, while an \textit{asynchronous} scheduler allows the CPU to perform other computations while the \gls{gpu} is active. The original GrCUDA scheduler is \textit{serial} and \textit{synchronous}, while our scheduler is \textit{parallel} and \textit{asynchronous}.

CUDA streams are key to enable parallel and asynchronous computation. In our scheduler, the allocation and management of streams are performed transparently by a stream manager. The stream manager also tracks what computations are currently active in each stream and handles events used for synchronization. Users can specify different policies to create new streams and to associate them with computations.
Existing streams are managed in FIFO order, and new streams are created only if no currently empty stream is available to schedule a given computation. If a computation has multiple children (i.e. computations that depend on it), the first child is scheduled on the parent's stream to minimize synchronization events, while following children are scheduled on other streams to guarantee concurrency. Simpler policies (e.g. schedule all children on a single stream) further reduce the scheduling costs; that said, our experiments always use the more general policies and show negligible scheduling overheads (\cref{sec:grcuda_cuda_speedup}).
The stream manager is architecture-aware: on \gls{gpu} architectures older than Pascal, the CPU cannot access \gls{um} if a kernel is active in the \gls{gpu}. The stream manager restricts each array's visibility to the stream where it is used until the CPU needs to access the array. The CPU is made temporarily unaware of the existence of arrays being used by the \gls{gpu}, and can access currently unused arrays even if the \gls{gpu} is active. While this optimization is not required on architectures since Pascal (thanks to its page fault mechanism),
our scheduler can automatically prefetch data to optimize transfers.

% Placed here for formatting
\begin{figure*}[t]
    \centering
    \includegraphics[width=\linewidth]{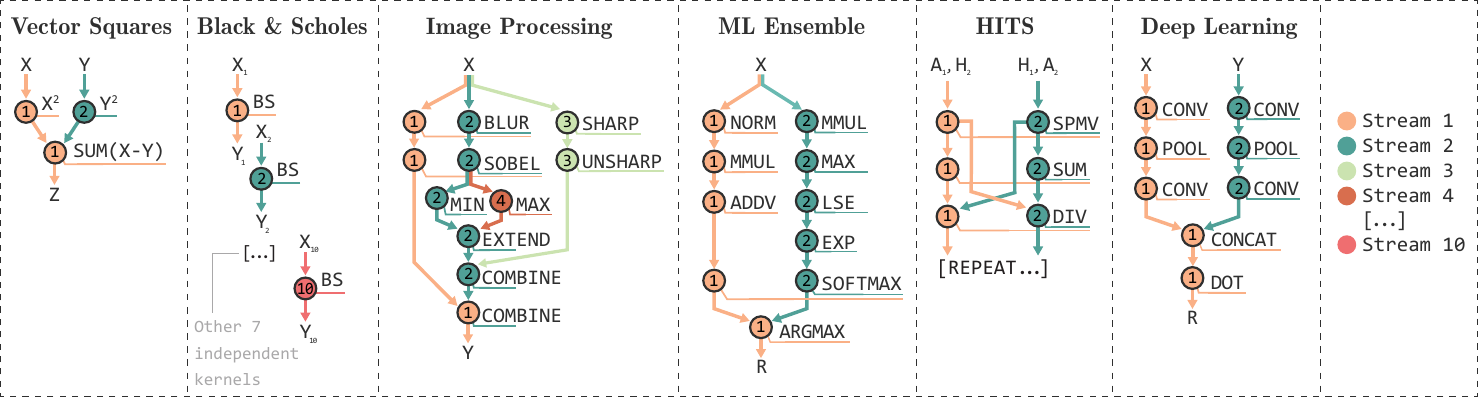}\\
    \caption{Computation structure of each benchmark, expressed as a \acrshort{dag} of \acrshort{gpu} kernel computations (denoted as circles). Colors denote CUDA streams, and kernels whose incoming arrows have different colors will require a synchronization event.}
    \label{fig:benchmark_structure}
\end{figure*}

\subsection{Language Design and Integration}\label{sec:language}

Our scheduler leverages the GrCUDA language's existing features and does not introduce any user-facing modification to the language. Kernel signatures are specified using \gls{nidl} or Truffle \gls{nfi}, simple typing systems that support basic data types and pointers.
% Arrays and arbitrary data structures are handled through untyped pointers, which does not limit the information that our scheduler can leverage for optimizations.
Optional argument annotations such as \texttt{input}, \texttt{output} or \texttt{const} are used by the scheduler to optimize computations that contain read-only arguments (\texttt{input} or \texttt{const}). For arguments without annotations, the scheduler treats them as modifiable by the kernel;
not specifying arguments as read-only does not affect correctness, but might limit the scheduler from performing further optimizations.

\section{Experimental Evaluation}\label{sec:experimental_results}

In this section, we evaluate the performance of our GrCUDA \gls{gpu} scheduler against multiple benchmarks that exhibit task-level parallelism and opportunities to leverage space-sharing and transfer-computation overlap to achieve lower execution time than a serial scheduler (\cref{sec:benchmarks}). First, we compare against the serial GrCUDA scheduler (\cref{sec:grcuda_speedup}) and show how our scheduler can exploit untapped \gls{gpu} resources to deliver better performance than a na\"ive serial scheduler.

Then, we compare the performance of the GrCUDA against the C++ CUDA API, measure how the GrCUDA scheduling is identical to the best hand-tuned scheduling possible, and how GrCUDA does not add any significant slowdown in the benchmark execution times (\cref{sec:grcuda_cuda_speedup}).
% We also estimate the best theoretical performance that can be achieved assuming that computations and data-transfer can always be overlapped regardless of hardware limitations (\cref{sec:theoretical_speedup}), and once again show how our scheduler is often very close to the intrinsic parallelism limit of each benchmark.
% We also estimate the amount of resource contention introduced by hardware space-sharing (\cref{sec:theoretical_speedup}), and once again show how our scheduler is often very close to the intrinsic parallelism limit of each benchmark.

Finally, we investigate, for each benchmark, the nature of the achieved speedup. First, we measure the amount of resource contention introduced by hardware space-sharing (\cref{sec:theoretical_speedup}); second, we measure what overlaps are present (transfer-computation or space-sharing), and we analyze how well each benchmark is using hardware resources such as device memory and L2 cache to understand which workloads are more suitable for asynchronous execution (\cref{sec:hardware_metrics}).

\subsection{Evaluation Setup}\label{sec:evaluation_setup}

Tests are performed on 3 different Nvidia GPUs with different architectures: a Tesla P100 (Pascal, 12 GB of device memory), a GTX 1660 Super (Turing, 6 GB), and a GTX 960 (Maxwell, 2 GB). \glspl{gpu} are connected to their respective host machines through \gls{pcie} 3.0. Testing consumer-grade \glspl{gpu} such as the GTX 1660 Super and the GTX 960 shows how our scheduler does not demand high-end data-center hardware to provide benefits, and it is useful for quick prototyping on commodity \glspl{gpu} and to accelerate desktop applications. 
% In any case, we see from \cref{sec:grcuda_speedup} how speedups are mostly constant with respect to input size, even when filling the \gls{gpu} memory.
Benchmarks are executed 30 times on random data.
% , removing the first iteration to obtain more consistent measurements.
% to provide time for the GraalVM host language (Python, in our case) to stabilize its performance through partial evaluation. 
We select the input sizes in each benchmark to use between 10\% and 90\% of the available memory on each \gls{gpu}, up to the largest size that fits in device memory.
% \glspl{gpu} are tested with different sizes as small values do not represent a reasonable use-case for data-center \glspl{gpu} like the P100, while large inputs do not fit on all \glspl{gpu}.
Execution time is the total amount of time spent by \gls{gpu} execution, from the first kernel scheduling until the end of execution. Parameters (e.g. the number of blocks) are optimized for best performance in serial execution to provide a worst-case comparison. The x-axes in \Cref{fig:grcuda_speedup,fig:grcuda_cuda_speedup,fig:theoretical_speedup} report the benchmark scale, a value proportional to the benchmark's memory footprint (e.g. the number of pixels in each input image).
% , instead of picking parameters for which the parallel scheduler performance is maximized or using policy-dependent settings.

\renewcommand\theadalign{tl}
\renewcommand\theadfont{\bfseries}
\setlength\tabcolsep{5pt}

\begin{table}
\centering
\ra{1.2}
    \caption{Amount of device memory for different input sizes in each benchmark. \glspl{gpu} are tested with different input sizes up to the largest size that fits in \gls{gpu} memory.
    % Input sizes are different on each \gls{gpu} and are chosen to provide similar percentage memory footprint.
    }
    
    \resizebox{\columnwidth}{!}{
	\begin{tabular}{@{}llll@{}}
		\toprule
		& \multicolumn{3}{c}{Memory footprint (GB)} \\
		\cmidrule{2-4}
		\thead{Benchmark name} & \thead{GTX 960} &  \thead{GTX 1660 Super} & \thead{Tesla P100} \\
		\midrule
		
		\textbf{Vector Squares (VEC)} & 0.4 GB - 1.9 GB & 0.4 GB - 3.1 GB & 0.4 GB - 11 GB \\
		\textbf{Black \& Scholes (B\&S)}  & 0.4 GB - 1.9 GB & 0.4 GB - 3.1 GB & 0.4 GB - 11 GB \\
		\textbf{Images (IMG)} & 0.2 GB - 1.0 GB  & 0.2 GB - 5.1 GB & 0.2 GB - 9.1 GB \\
		\textbf{ML Ensemble (ML)}  & 0.4 GB - 1.9 GB & 0.4 GB - 3.3 GB & 0.4 GB - 9.9 GB \\
		\textbf{HITS} & 0.4 GB - 1.5 GB & 0.4 GB - 4.2 GB & 0.4 GB - 9.9 GB \\
		\textbf{Deep Learning (DL)} & 0.3 GB - 1.4 GB & 0.3 GB - 4.9 GB & 0.3 GB - 6.5 GB \\
		
	    \midrule
	    \textbf{GPU device memory} & 2 GB & 6 GB & 12.2 GB \\
	    
		\bottomrule
	\end{tabular}
    }
    \label{tab:benchmark}
\end{table}

\subsection{Benchmark Suite}\label{sec:benchmarks}

We tested our scheduler on 6 benchmarks and a total of 33 different kernels representing common \gls{gpu} workloads (image processing, machine learning, etc.) and containing opportunities for task-level parallelism through space-sharing and computation-transfer overlap. To the best of our knowledge, no existing \gls{gpu} benchmark suite has the goal of evaluating intra-application task-level parallelism, as most benchmark suites (e.g. Rodinia \cite{che2009rodinia}) focus on single kernels and sequential execution.
Still, we take or derive the CUDA kernels in our benchmarks from open-source implementations.

We scale the input size linearly to visualize more clearly if any hardware bottleneck impacts performance as input size exceeds a threshold.
For instance, we change the number of rows for the matrix multiplications in the ML benchmark, but keep fixed the number of features and output classes. 

\Cref{fig:benchmark_structure} presents each benchmark's task dependency structure and highlights the optimal stream assignment for each kernel. \Cref{tab:benchmark} summarizes each benchmark. 
% \textit{Input size} is the number of data elements (e.g. the size of an array or values in a matrix) without additional data structures such as output arrays, whose size can depend on the input.
% \textit{Memory footprint} is the percentage of \gls{gpu} memory used by each benchmark. 
% We chose the input sizes so that the \textit{memory footprint} of each benchmark ranges approximately between 10\% and 90\% of the total \gls{gpu} memory.
The chosen input sizes guarantee that the \textit{memory footprint} covers both small and large computations compared to the total memory of each \gls{gpu}.
% (including output and other data structures).
For each benchmark, we present a brief description.

\begin{itemize}
    \item \textbf{Vector Squares (VEC)}: a simple benchmark that measures a basic case of task-level parallelism and computes the sum of differences of 2 squared vectors. Each iteration has new input data, simulating a streaming computation that requires transfer from CPU to \gls{gpu}.
    Inspired by \cite{reductionkepler}.
    
    \item\textbf{Black \& Scholes (B\&S)}: Black \& Scholes equation for European call options, for 10 underlying stocks, and 10 vectors of prices. Adapted from \cite{bs} to simulate a computationally intensive streaming benchmark with double-precision arithmetic and many independent kernels that can be overlapped with no dependencies. 
    
    \item \textbf{Image Processing (IMG)}: an image processing pipeline that combines a sharpened picture with copies blurred at low and medium frequencies \cite{blur}, to sharpen the edges, soften everything else, and enhance the subject. 
    % for portrait retouching, where a photographer desires to enhance facial features while smoothing the subject's skin and the background.
    The benchmark has complex dependencies on 4 streams.
    
    \item \textbf{Machine Learning Ensemble (ML)}: an ML pipeline that combines Categorical Na\"ive Bayes and Ridge Regression classifiers by applying softmax normalization and averaging scores. The input matrix has 200 features. This benchmark contains branch imbalance (the Na\"ive Bayes classifiers takes longer) and read-only arguments.
    
    \item \textbf{HITS}: it computes the HITS algorithm on a graph \cite{kleinberg1999authoritative} using repeated \gls{spmv} on a matrix and its transpose, and is implemented with LightSpMV \cite{liu2015lightspmv}. It contains complex cross-synchronizations and multiple iterations.
    % The input graph has degree 3 and uniform distribution.

    \item \textbf{Deep Learning (DL)}: a convolutional neural network that projects 2 input images to low dimensional embeddings and combines the embeddings using a dense layer. Similar neural networks can be used, for example, to classify if 2 images contain the same subject.
\end{itemize}

\subsection{Performance against Serial GrCUDA Scheduling}\label{sec:grcuda_speedup}

\begin{figure}[t]
    \centering
    \includegraphics[width=\columnwidth]{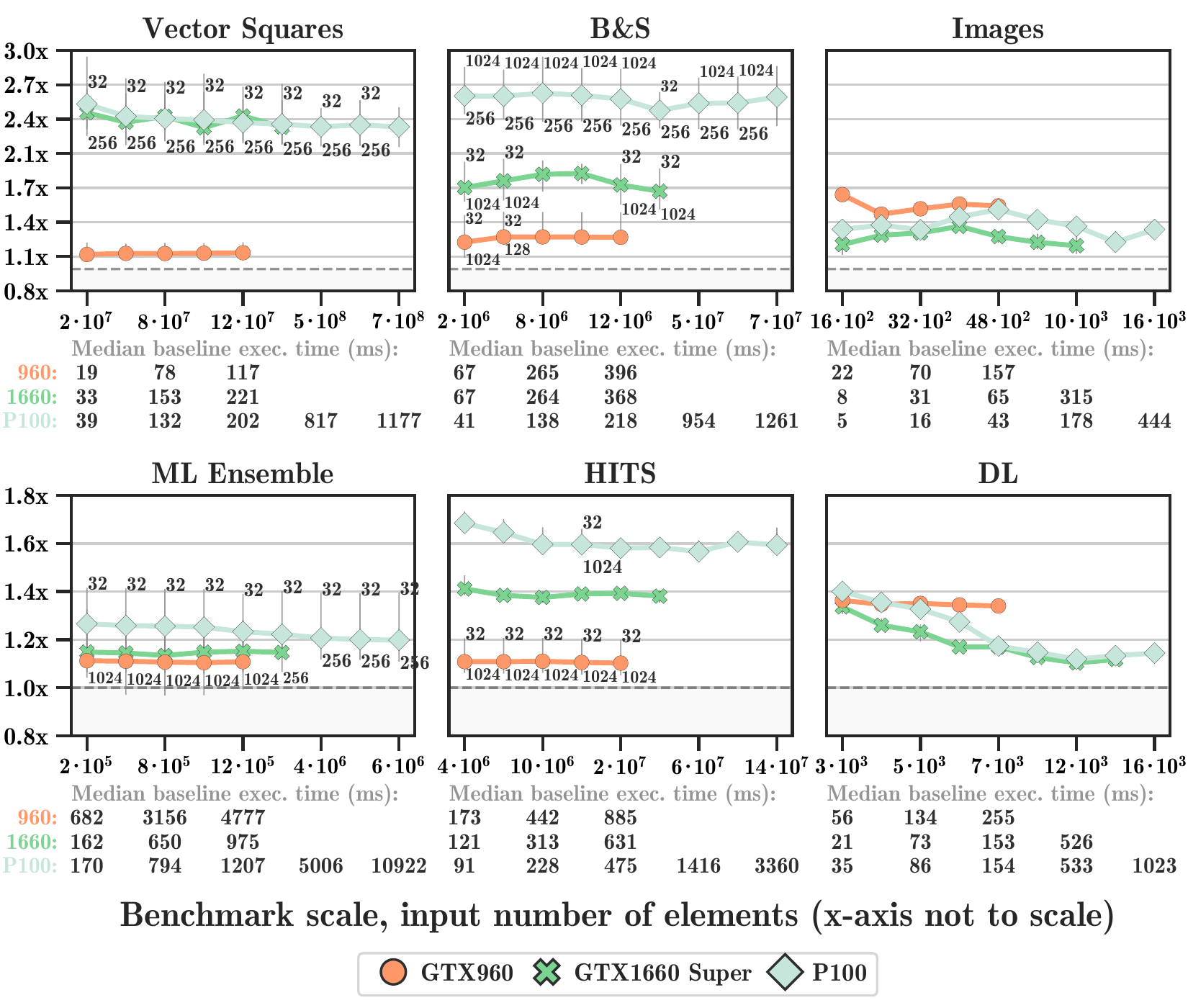}\\
    \caption{Parallel scheduler speedup over serial schedugler. Our parallel scheduler provides a geomean speedup of 44\% over the original GrCUDA serial scheduler. We highlight block sizes giving the best/worst speedup (when significant).
    % X-axis not to scale.
    }
    \label{fig:grcuda_speedup}
\end{figure}

We compare the performance of our parallel scheduler against the GrCUDA serial scheduler (\Cref{fig:grcuda_speedup}).
Automatic data prefetching is enabled on the Tesla P100 and the GTX 1660 Super, while on the GTX 960, data is necessarily transferred ahead of the computation as it does not have a page fault mechanism.
We always deliver better performance over the serial scheduler, with a geomean speedup of 44\% on the 3 \glspl{gpu}. 
The GTX 960 is 25\% faster, while the P100 performs the best, with a geomean speedup of 61\%. More hardware resources, together with automatic prefetching, results in better parallelization, and we show how our approach works out-of-the-box on data-center \glspl{gpu}.
While still faster than the serial baseline, disabling automatic prefetching is not recommended: concurrent kernel execution turns the page fault controller into the main bottleneck, limiting the benefits of overlapping data transfer with computation. 
% In \cref{fig:grcuda_speedup}, \textit{serial scheduling} measures synchronous serial \gls{gpu} execution. In this case, dependencies between kernels are not computed, making GrCUDA overheads even smaller. \textit{Parallel scheduling} (or DAG scheduling) is the computation time when using our GrCUDA asynchronous parallel scheduler and performing \gls{gpu} resource-sharing.
When using \textit{serial scheduling}, GrCUDA does not compute dependencies, making overheads even smaller.
We average results for \textit{block sizes} (the number of threads in each 1D block) from 32 to 1024.
% A higher number implies bigger blocks, and possibly fewer concurrent blocks running in parallel.
In benchmarks with 2D and 3D blocks (e.g. IMG and DL), we keep 2D blocks with size 8x8 and 3D blocks with size 4x4x4, as bigger blocks resulted in longer execution times in every case. 

Speedups are mostly independent of the input data size, as we sweep through inputs with size from less than 10\% to almost 100\% of the available \gls{gpu} memory. 
Even if kernels fill the \gls{gpu} resources, it is still possible to achieve a speedup by overlapping data transfer with other kernels' execution. 
% Our DAG scheduler is never worse than the serial scheduler: users can always leverage our scheduler instead of profiling their code to find the best policy.

DAG scheduling appears to be more robust to different kernel configurations: in many cases (such as VEC and HITS), using \texttt{block\_size=32} results in higher speedup, but similar execution time as with larger block size. With serial scheduling, small blocks result in under-utilization of \gls{gpu} resources such as shared memory, while DAG scheduling provides better utilization by having multiple kernels run in parallel. Thanks to DAG scheduling, programmers have to spend less time profiling their code to find the optimal kernel configuration.

\begin{figure}[t]
    \centering
    \includegraphics[width=\columnwidth]{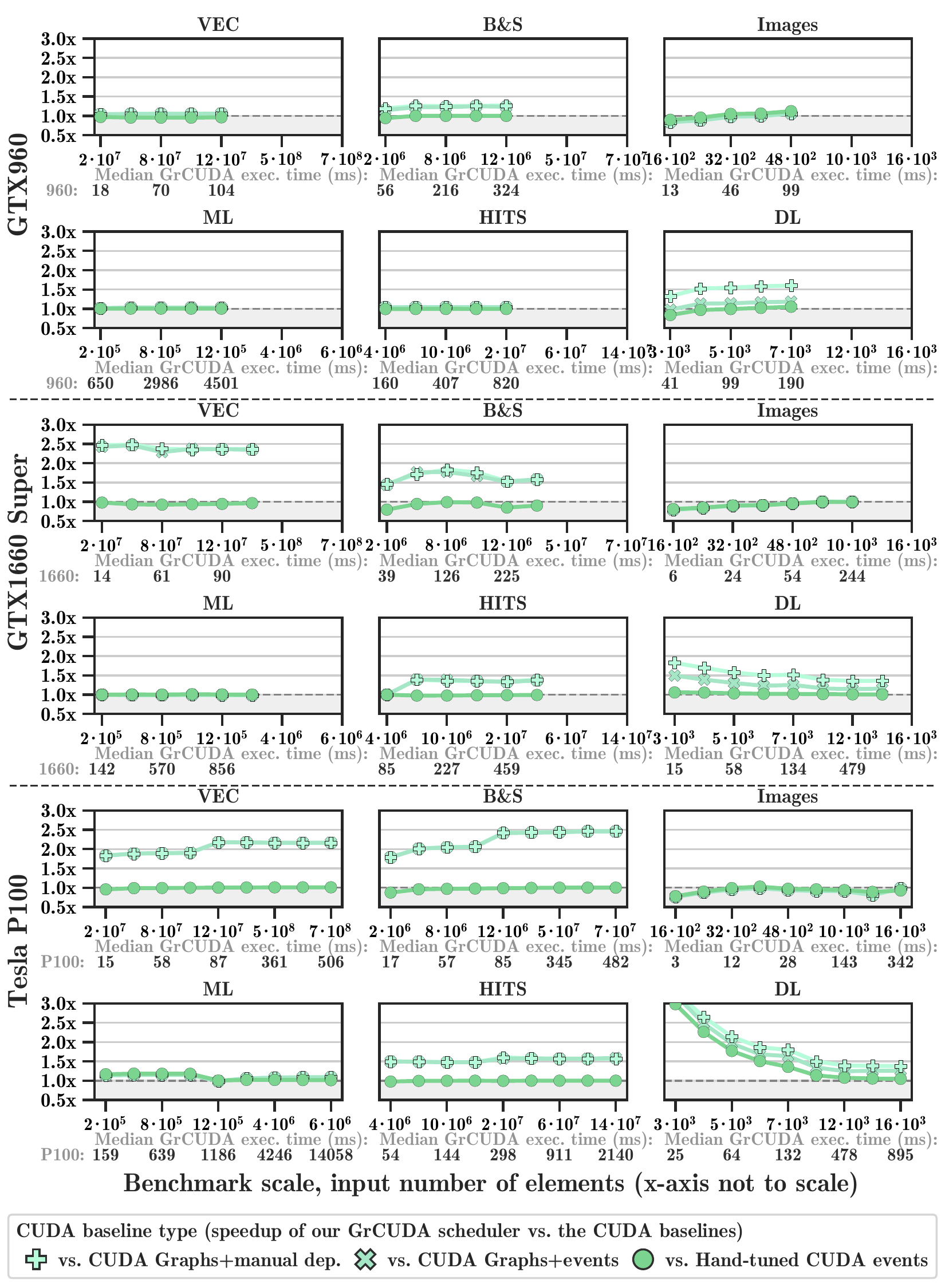}\\
    \caption{Speedup of our GrCUDA scheduling
against hand-optimized CUDA Graphs
(higher is better). We are always faster, and the GrCUDA runtime overheads are negligible for computations lasting more than a few milliseconds.
    %  X-axis not to scale.
    }
    \label{fig:grcuda_cuda_speedup}
\end{figure}

\subsection{Performance against CUDA Graphs}\label{sec:grcuda_cuda_speedup}

To understand how our GrCUDA scheduler performs against existing solutions, we re-implemented our benchmarks using the C++ CUDA Graphs API. The kernel code and the setup (e.g. input and block sizes) are the same as GrCUDA, but the host code is written using the C++ CUDA Graphs API.
We test CUDA Graphs using \textit{stream-capture} to wrap hand-optimized multi-stream scheduling synchronized with CUDA events and using the Graph API to specify dependencies between computations manually.
These CUDA Graphs are built only once per execution, and overheads are completely amortized over many iterations.
% We also experiment with CUDA Graphs on a single stream to measure if GrCUDA introduces significant overheads through its GraalVM runtime.
Finally, we provide a hand-optimized implementation purely based on CUDA events to have full control over data movement and simulate CUDA Graph's performance if it supported data prefetching. 

Our GrCUDA scheduler, in addition to being fully automated, is never significantly slower than any of the CUDA Graphs baselines and is often faster (\Cref{fig:grcuda_cuda_speedup}). The large performance gaps compared to CUDA Graphs seen on the GTX 1660 Super and the P100 are mostly explained by our automatic prefetching, which the CUDA Graphs API seems unable to perform.
Even when enabling prefetching in the CUDA baseline, our parallel scheduler achieves equal performance to the hand-optimized baseline.
Execution time speedups are hardly affected by input size, with minor differences only in computations lasting a few milliseconds and containing many dependencies (such as IMG).

% In the case of asynchronous kernel execution, synchronization events have been optimized by hand to minimize execution time and resource usage. Serial scheduling uses the default stream, with kernels running one after another without overlaps.
% This procedure measures how large the overhead introduced by GrCUDA is (especially the automatic computation of kernel dependencies) and how far our GrCUDA scheduler is from the optimum in terms of how many synchronization points and streams are created. 
% The gap between CUDA and GrCUDA is below 2\% on average (\cref{fig:grcuda_cuda_speedup}), and converges to 0 as the input size increases.
% Overhead are noticeable only for minuscule computations (a few milliseconds) and contain a large number of dependencies, as in IMG with an $1600 {\times} 1600$ input image: using GrCUDA in any realistic computation will not decrease performance, regardless of the scheduling policy. 

\subsection{Impact of Space-Sharing Resource Contention}\label{sec:theoretical_speedup}

\begin{figure}[t]
    \centering
    \includegraphics[width=\columnwidth]{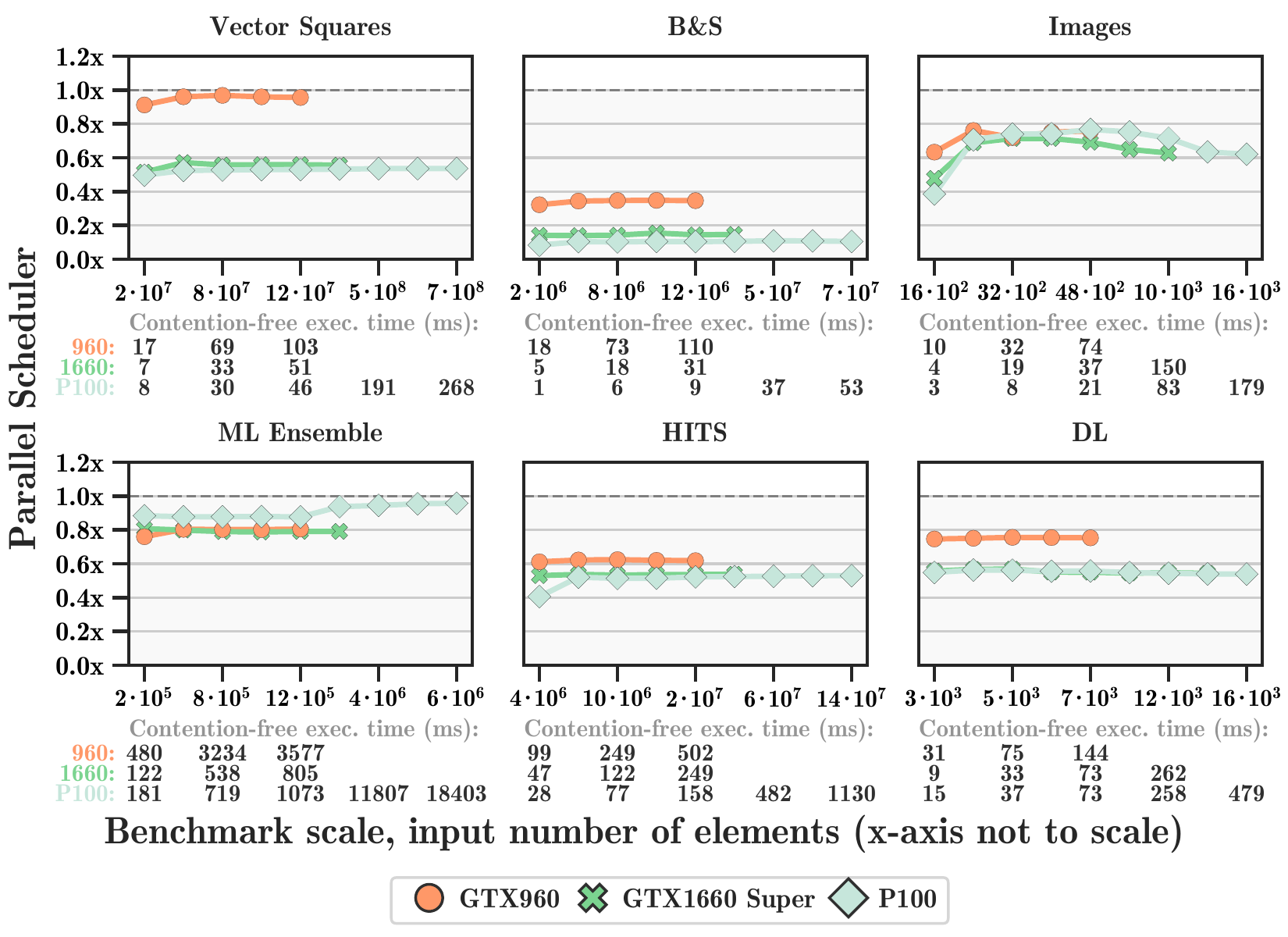}\\
    \caption{
    % In most cases, resource contention introduced by hardware space-sharing has a small impact on the execution time of our benchmarks, with a performance loss below 30\%.
    Slowdown compared to execution
without hardware resource contention. Space-sharing introduces a performance loss of around 30-40\% due to hardware space-sharing.
    }
    \label{fig:theoretical_speedup}
\end{figure}

By looking at dependencies between kernels and measuring their execution time with serial scheduling so that each kernel has full access to the \gls{gpu} resources, we estimate the resource contention on the \gls{gpu} hardware and the \gls{pcie} bandwidth introduced by space-sharing. 
% We assume a GPU that can compute multiple kernels in parallel without any resource contention and obtain the contention-free execution time for each benchmark.
\Cref{fig:theoretical_speedup} shows how far each benchmark is from its theoretical contention-free peak performance.
% Results refer to the GTX 960, a middle-range consumer-grade GPU where resource contention is the most visible. 
% Through this analysis, we understand the impact that resource contention has on each benchmark, and possibly how much faster each benchmark could be on a more powerful \gls{gpu} in which resource contention is less visible.
Results are mostly consistent between \glspl{gpu}, with a relative execution time that is often around 70\% of the contention-free performance bound; while resource contention is present, it is small enough to make space-sharing worthwhile.
Unsurprisingly, B\&S, which is composed of 10 independent computations, achieves around 15-20\% of its contention-free peak performance due to limitations on \gls{pcie} bandwidth and double-precision arithmetic units availability. 
% running all of them in parallel would require an exceedingly large number of double-precision arithmetic units.
% , although our parallel scheduler is still faster than serial execution.

\subsection{Analysis of Hardware Utilization}\label{sec:hardware_metrics}

It is fascinating to understand, for each benchmark, the nature of the speedup achieved through resource sharing, i.e., if the speedup is caused by overlapping computation with data transfer or if the speedup is explained by higher utilization of \gls{gpu} resources such as device memory bandwidth through space-sharing. First, we measure how much overlap is present in each benchmark. We measure 4 different types of overlap:
\begin{itemize}
    \item \textbf{CT}, computation against transfer: percentage of \gls{gpu} kernel computation that overlaps with any data transfer
    %(host-to-device or vice-versa)
    \item \textbf{TC}, transfer against computation: percentage of data transfer that overlaps with any kernel computation(s)
    \item \textbf{CC}, percentage of \gls{gpu} computation overlapped with other \gls{gpu} computation
    \item \textbf{TOT}, any type of overlap: here we consider any type of overlap between data-transfer and/or computations.
    If a computation/data-transfer overlaps more than one computation/data-transfer, the overlap is counted only once (we consider the union of the overlap intervals)
\end{itemize}

\begin{figure}[t]
    \centering
    \includegraphics[width=\columnwidth]{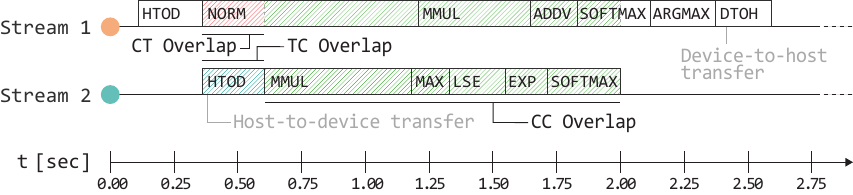}\\
    \caption{Example of a possible execution timeline for the ML benchmark. We highlight different types of overlap between transfer and computation, as defined in \cref{sec:hardware_metrics}.}
    \label{fig:timeline}
\end{figure}

\Cref{fig:timeline} shows how we compute these overlaps starting from the execution timeline.
% Measures are taken for the largest input sizes in the evaluation.
% , for the block size that results in higher speedup, to obtain a clearer understanding of what type of overlap is providing the speedup.
Although the TOT overlap can be a good proxy of the achieved speedup, it is sometimes inflated by high CC overlap, as overlapping computations does not always translate to faster execution.
% , especially if individual kernels have enough blocks to fill the GPU processors.
In VEC, the speedup comes only from transfer and computation overlap, while the overlap of kernels that leave a large amount of shared memory unused if executed serially explains the speedup in IMG.
% Very different values of CT and TC (as in \textbf{B\&S}) indicate that the computation lasts much longer than the data transfer, and part of the computation is not overlapped: a faster computation would result in higher CT overlap and better speedup.

Computation time and transfer time do not increase with the same proportionality factor as the input size; small input data do not use the \gls{pcie} bandwidth fully, and a 10x increase in data size might translate into a transfer time increase below 10x, up to the available bandwidth.
On faster \glspl{gpu} the computation time is lower, while the transfer time is roughly identical to less powerful hardware (assuming the same transfer interface): more computation is overlapped to data transfer, leading to better speedups. For example, in the B\&S benchmark, the CT overlap increases on faster \glspl{gpu}, and so does the speedup.
% We expect that a faster \gls{gpu} provides speedups in line or better than what reported here: on a faster \gls{gpu} the computation time is lower, while the transfer time is roughly identical (assuming the same transfer interface); as such, the percentage of computation overlapped to data transfer is higher, leading to even better speedups.

\begin{figure}[t]
    \centering
    \includegraphics[width=\columnwidth]{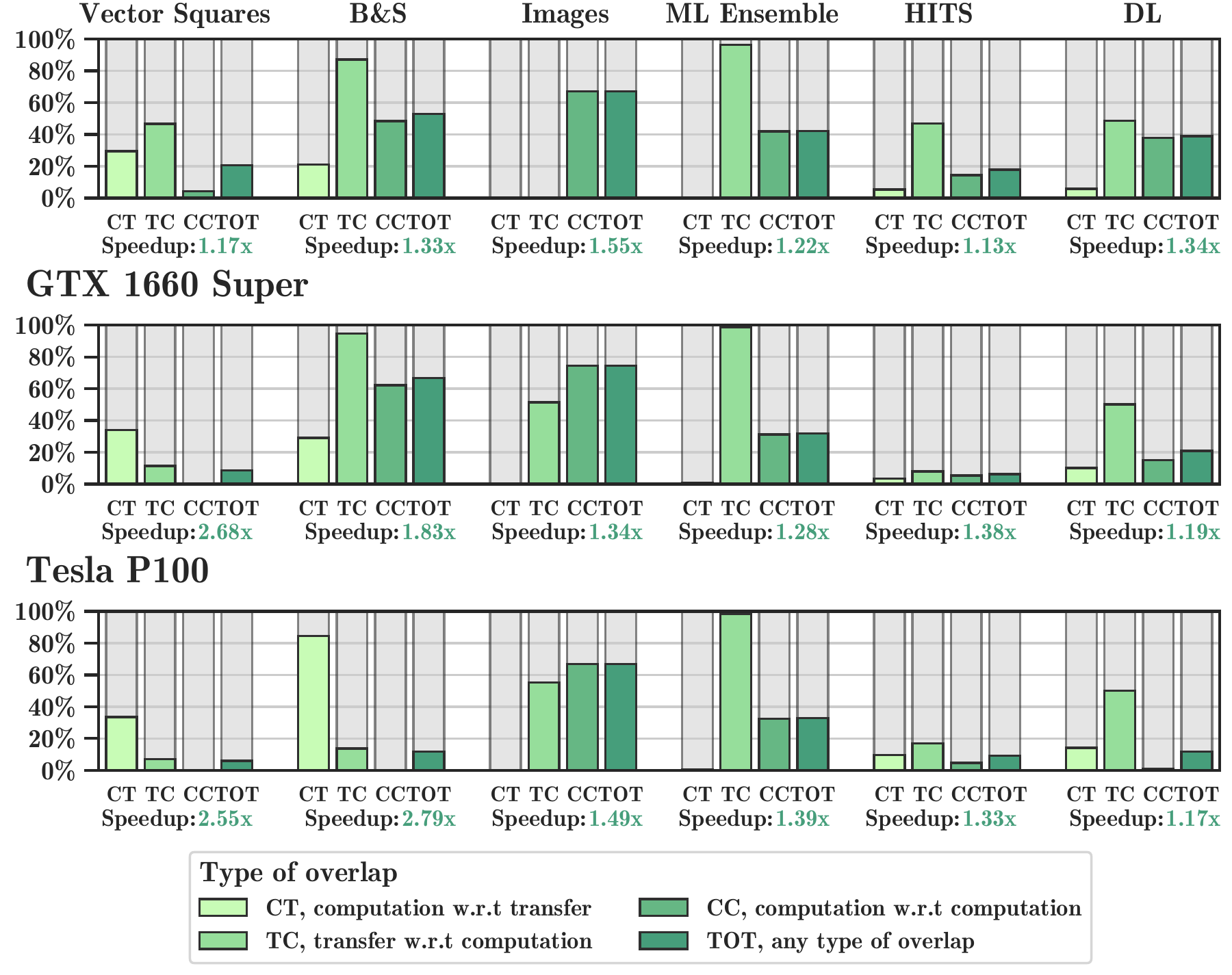}\\
    \caption{
    Amount of transfer and computation overlap for each benchmark, for serial and parallel scheduling. We report below each plot the speedup obtained in the benchmark.}
    \label{fig:overlap}
\end{figure}

\begin{figure}[t]
    \centering
    \includegraphics[width=0.96\columnwidth]{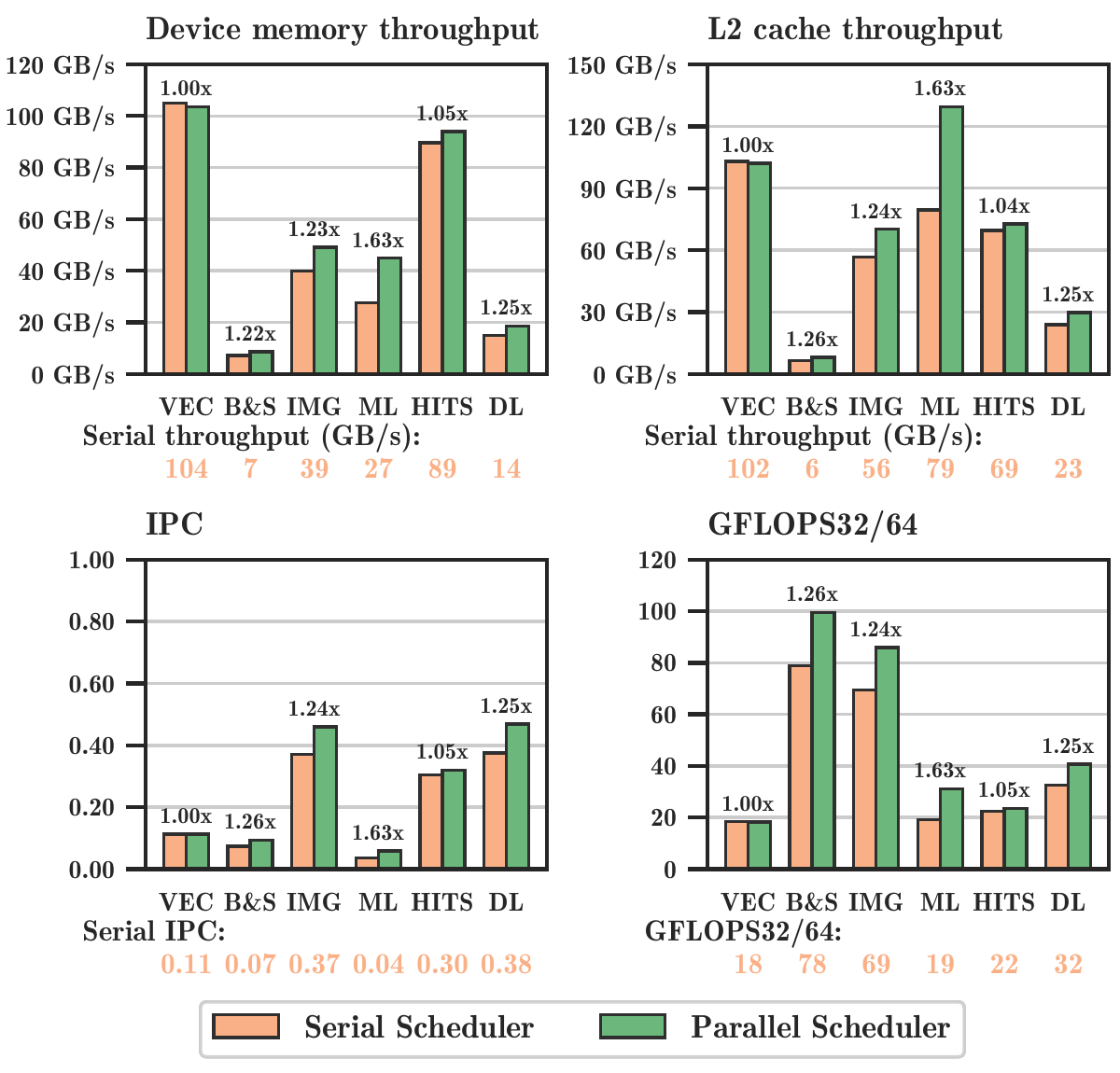}\\
    \caption{
    Hardware metrics for each benchmark and execution policy, as measured on the GTX 1660 Super.
    All benchmarks in which different kernels overlap their execution show an increase in hardware utilization.}
    \label{fig:memory}
\end{figure}

We then analyze how our parallel scheduler affects hardware-level metrics such as device memory throughput, L2 cache throughput, \gls{ipc}, and GFLOPS. We use \texttt{nvprof} and \texttt{ncu}\footnote{\url{https://docs.nvidia.com/cuda/profiler-users-guide/index.html}, \url{https://docs.nvidia.com/nsight-compute/NsightComputeCli/index.html}} to measure the number of bytes read/written by each kernel from/to device memory and L2 cache, and the total number of instructions executed.
As collecting these metrics introduces high execution time overhead and prevents the execution of concurrent kernels, we combine the execution timeline without metric collection with hardware metrics collected in separate runs (variance in these metrics between different runs is insignificant).
Metrics are collected only on the GTX 1660 Super as we did not have root-level access to the Tesla P100.
The amount of bytes read/written and the total number of instructions executed \textit{by each kernel} mostly depends on the kernel itself and is not significantly impacted by space-sharing; as such, this evaluation is useful to estimate the global \gls{gpu} behavior when space-sharing is performed. GFLOPS is estimated from the total number of floating-point operations (single and double precision).

\Cref{fig:memory} shows how in kernels with computation overlap (e.g. ML and IMG), the increase in memory throughput is significant and in-line with the total speedup observed for these benchmarks. VEC does not have any increase in memory throughput, as its speedup comes exclusively from transfer overlap.
% Similarly, \textbf{B\&S} shows a small memory throughput increase, as the benchmark main bottleneck is the arithmetic intensity (note its high GFLOPS value), and the biggest part of the speedup comes from transfer overlap.
Benchmarks that operate on dense matrices make heavier use of L2 cache, whose throughput increases with parallel scheduling. 
The low \gls{ipc} in ML is caused by a slow kernel that operates on tall matrices and does not use the \gls{gpu} parallelism to its full extent: running multiple kernels in parallel hides its latency and provides the speedup in \Cref{fig:grcuda_speedup}. 

From these analyses, we understand what limits performance in each benchmark.
% For example, B\&S has high TC and low CT on the GTX 1660 (\cref{fig:overlap}): the computation lasts longer than the data transfer, and part of the computation is not overlapped; on the other hand, on the Tesla P100 the computation is completely masked by transfer (high CT), and indeed we observe a better speedup. 
% As B\&S performs complex mathematical operations on independent values, it shows very high GFLOPS count (\cref{fig:memory}) and almost no cache utilization.
% Faster arithmetic units would yield even higher speedups: the transfer time would not be affected, resulting in higher overlap with data-transfer.
% With faster arithmetic units, the overlap with data-transfer would increase (as the transfer time is unaffected), and the speedup w.r.t serial execution would be even higher.
For example, B\&S performs complex mathematical operations on independent values, with a very high GFLOPS count (\Cref{fig:memory}) and almost no cache utilization. On the GTX 1660, B\&S
has high TC and low CT (\Cref{fig:overlap}); the computation lasts longer than the data transfer, and part of the computation is not overlapped; on the other hand, the Tesla P100, which has 20x higher double-precision performance than the 1660, completely masks the computation with transfer (high CT), and indeed we observe a better speedup. 
Improving performance even further requires lowering the transfer time, for example, through \gls{pcie} 4.0.

\section{Conclusion and future work}\label{sec:conclusion}

We presented a novel scheduler for \gls{gpu} computations that can automatically infer data dependencies to build a computation \gls{dag} at run time. The scheduler allows computations to execute in parallel through \gls{gpu} space-sharing and overlaps data-transfer and execution whenever possible.

We validate our scheduler on 6 benchmarks and a total of 33 \gls{gpu} kernels. Our scheduler provides a geomean speedup of 44\% (up to 270\%) over serial synchronous scheduling, and is always faster. It automatically achieves the same scheduling as hand-optimized CUDA Graphs code, without any slowdown.

Our scheduler seamlessly integrates with the GrCUDA environment, a polyglot CUDA API based on GraalVM that provides easy access to \gls{gpu} acceleration to languages such as Java, Python, JavaScript, and R.
Users can leverage our work without knowledge of the underlying scheduler, and without changing their code. It is not required to specify in advance the code structure or control flow, as our scheduler dynamically executes computations as provided by the host program. 
Our work simplifies \gls{gpu} code prototyping and acceleration by giving easy access to untapped \gls{gpu} resources at no cost, while masking the complexity of asynchronous \gls{gpu} computations.

As future work, we plan to extend our technique to multiple \glspl{gpu}: the problem is significantly harder, as it requires to compute data location and migration costs at run time to identify the optimal scheduling. We will also leverage run time information for additional optimizations, such as estimating the ideal block size based on data size and previous executions.

\section*{Acknowledgments}
We thank Oracle Labs for its support and contributions to this work. The authors from Politecnico di Milano are funded in part by a research grant from Oracle. We also thank Rene Mueller and Lukas Stadler, the original authors of GrCUDA, for their valuable feedback and opinions.
Oracle and Java are registered trademarks
of Oracle and/or its affiliates.
% Other names may be trademarks of their respective owners.

\bibliographystyle{IEEEtran}
% Balance columns, put a column break after reference 22
% \IEEEtriggeratref{23}

\bibliography{references}

\end{document}